\documentclass[aps, notitlepage, onecolumn, prb, superscriptaddress]{revtex4-1}
\usepackage{hyperref}
\usepackage{amsmath}
\usepackage{amsfonts}
\usepackage{amssymb}
\usepackage{graphicx}
\usepackage{epstopdf}
\usepackage{color}

\usepackage{physics}

\newcommand{\As}{$^{75}$As }

\begin{document}

\title{Dynamic Polarization and Relaxation of $^{75}$As Nuclei in Silicon at High Magnetic Field and Low Temperature
}

\author{J. J\"{a}rvinen}
\email{jaanja@utu.fi}
\author{J. Ahokas}
\author{S. Sheludyakov}
\author{O. Vainio}
\author{L. Lehtonen}
\author{S. Vasiliev}
\affiliation{Wihuri Physical Laboratory, Department of Physics and Astronomy, University of Turku, 20014 Turku, Finland}
\author{D. Zvezdov}
\affiliation{Wihuri Physical Laboratory, Department of Physics and Astronomy, University of Turku, 20014 Turku, Finland}
\affiliation{Institute of Physics, Kazan Federal University, Russia}
\author{L. Vlasenko}
\affiliation{A. F. Ioffe Physico-Technical Institute, Russian Academy of Sciences, 194021 St. Petersburg, Russia}


\begin{abstract}

We present the results of experiments on dynamic nuclear polarization and relaxation of $^{75}$As in silicon crystals. Experiments are performed in strong magnetic fields of 4.6 T and temperatures below 1 K. At these conditions donor electron spins are fully polarized, and the allowed and forbidden ESR transitions are well resolved. We demonstrate effective nuclear polarization of $^{75}$As nuclei via the Overhauser effect on the time scale of several hundred seconds. Excitation of the forbidden transitions leads to a polarization through the solid effect. The relaxation rate of donor nuclei has strong temperature dependence characteristic of Orbach process.

\end{abstract}

\maketitle

\section{Introduction}
\label{intro}
Dynamic nuclear polarization (DNP) is a method for increasing nuclear polarization and enhancing the NMR signal in electron-nuclear spin systems. The recent interest in DNP has been raised by the possibility to utilize spins as qubits in quantum computers. Intensive studies in this field have been performed in systems including liquid state NMR,\citep{Cory1997}, nitrogen vacancy centers in diamond\citep{Fuchs2011} and shallow donors in semiconductors\citep{Kane1998}. Nuclear and electron spins in silicon have proven to have long coherence times \citep{Steger2012, Tyryshkin2012, Wolfowicz2015} and they can be controlled by means of magnetic resonance, \citep{Pla2013} optics, \citep{Lo2015} electric fields, \citep{Pica2014, Scheuer2016} and strain.\citep{Franke2015} DNP could be used as one of the tools to initialize and control the qubits in QC. Silicon has the advantage of being the material most heavily used by the semiconductor industry. Possible applications could be then easier to interlink with the present semiconducting technologies.

There are two main mechanisms of DNP which can be effectively utilized at high magnetic fields and low temperatures:\citep{Abragam1985} The Overhauser effect (OE) and the Solid effect (SE). In OE usually the allowed electronic transitions are saturated and the nuclear polarization accumulates due to simultaneous electron and nuclear (forbidden) spin flips caused by the thermal relaxation. In solid effect the forbidden transition is directly excited with microwaves, which is followed by electron spin relaxation via the allowed transition. The allowed and forbidden transitions are well resolved at sufficiently high magnetic fields, and thus it becomes possible to separate the OE and SE DNP. However, the microwave excitation probability as well as the thermal relaxation rate of forbidden transitions decreases in strong magnetic fields, which may impose certain limitations on the efficiency of the above mentioned methods. In our recent work\citep{Jarvinen2014} we have demonstrated that effective DNP of $^{31}$P in silicon can be performed at strong magnetic field, and very large nuclear polarization was achieved.  

In the previous experiments with $^{31}$P in silicon the OE DNP and relaxation of the nuclei was found to be bi-exponential.\citep{Jarvinen2014} We suggested that this could be caused by the nearest $^{29}$Si spins enhancing the flip-flop process of donor nucleus. As it is shown in this work, the fast part in the beginning of the $^{75}$As relaxation is absent, and the process is well fitted with a single exponent. This leaves a possibility that some other mechanism may be responsible for the double exponential time evolution in $^{31}$P.

\section{Experimental details}
\label{sec:1}
The sample was $6\times 6\times 0.4$ mm piece of commercial crystalline silicon doped with $\approx 6 \times 10^{16}$ cm$^{-3}$ of $^{75}$As and having $[001]$ crystal axis aligned with $B_0$ field. The sample was mounted on a flat mirror of a semi-confocal 130 GHz Fabry-Perot resonator (FPR). The upper spherical Cu mirror is connected to a mm-wave block of the cryogenic ESR spectrometer.\citep{Vasilyev2004} The doped silicon sample in the FPR destroys its resonances at room temperature  because of strong mm-power absorption by the semiconducting sample. The cavity resonances are recovered at low temperatures when the sample becomes dielectric. However, due to the high dielectric constant of silicon, the resonance frequencies are shifted down by $\approx$15 GHz from that of the empty cavity. This creates difficulties of tuning the cavity mode into the operating band of the ESR spectrometer. The problem was solved by mounting the lower mirror of the FPR to a piezo positioner\citep{Attocube}, capable of changing the separation between the mirrors by 5 mm at low temperatures. This is sufficient for selecting one of four different FPR modes within the tuning range of the spectrometer. The resonator with the sample is cooled by a dilution refrigerator down to a minimum temperature of $\approx 200$ $m$K. The ESR spectrometer operates in continuous wave mode without field modulation. The excitation power of the spectrometer can be tuned by more than seven orders of magnitude. The power used for detecting ESR signal is about 1 pW and maximum available excitation power used for DNP is about 20 $\mu$W.

\section{DNP of $^{75}$As in Silicon}
\label{sec:2}

\begin{figure}
\begin{center}
  \includegraphics[width=0.5\textwidth]{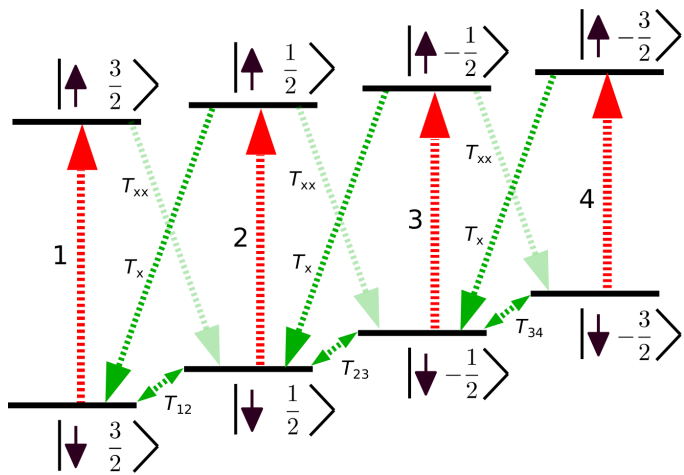}
\end{center}
\caption{Energy level diagram of $^{75}$As electron and nuclear spin states in strong magnetic field. The arrows mark the electronic spin state and the half integer numbers the As nuclear spin states. The long red arrows numbered from 1 to 4 indicate allowed ESR transitions and the long green arrows thermal flip-flop  $T_x$ and flip-flip $T_{xx}$ relaxation transitions. Double headed small arrows indicate the various nuclear relaxation paths.
}
\label{fig:Diag}       
\end{figure}

Arsenic (together with phosphorus) belongs to Mendeleev group V in the periodic table. In silicon crystal it becomes a donor of an electron with the energy level just about 50 meV below the conduction band minimum.\citep{Feher1959, Pica2014} The almost pure s-state wavefunction of the electron has Bohr radius of about 2 nm. The spin Hamiltonian can be written in the form:
\begin{equation}\label{hamiltonian}
H/ \hbar=-\gamma_e S_z B_0 + \gamma_n I_z B_0 +A S\cdot I,
\end{equation}
where $\gamma_e$ and $\gamma_n$ are electron and nuclear gyromagnetic ratios and $A=198.35$ MHz is \As  hyperfine constant.\citep{Feher1959} Quadrupolar frequency shift is very weak for As  and is not taken into account.\citep{Franke2015} The  solution of eq. (\ref{hamiltonian}) for spin 3/2 nucleus gives 8 energy eigenvalues which results in 4 allowed and 6 forbidden ESR transitions. These are schematically plotted in fig. \ref{fig:Diag}. 
\begin{figure}
\includegraphics[width=0.96\textwidth]{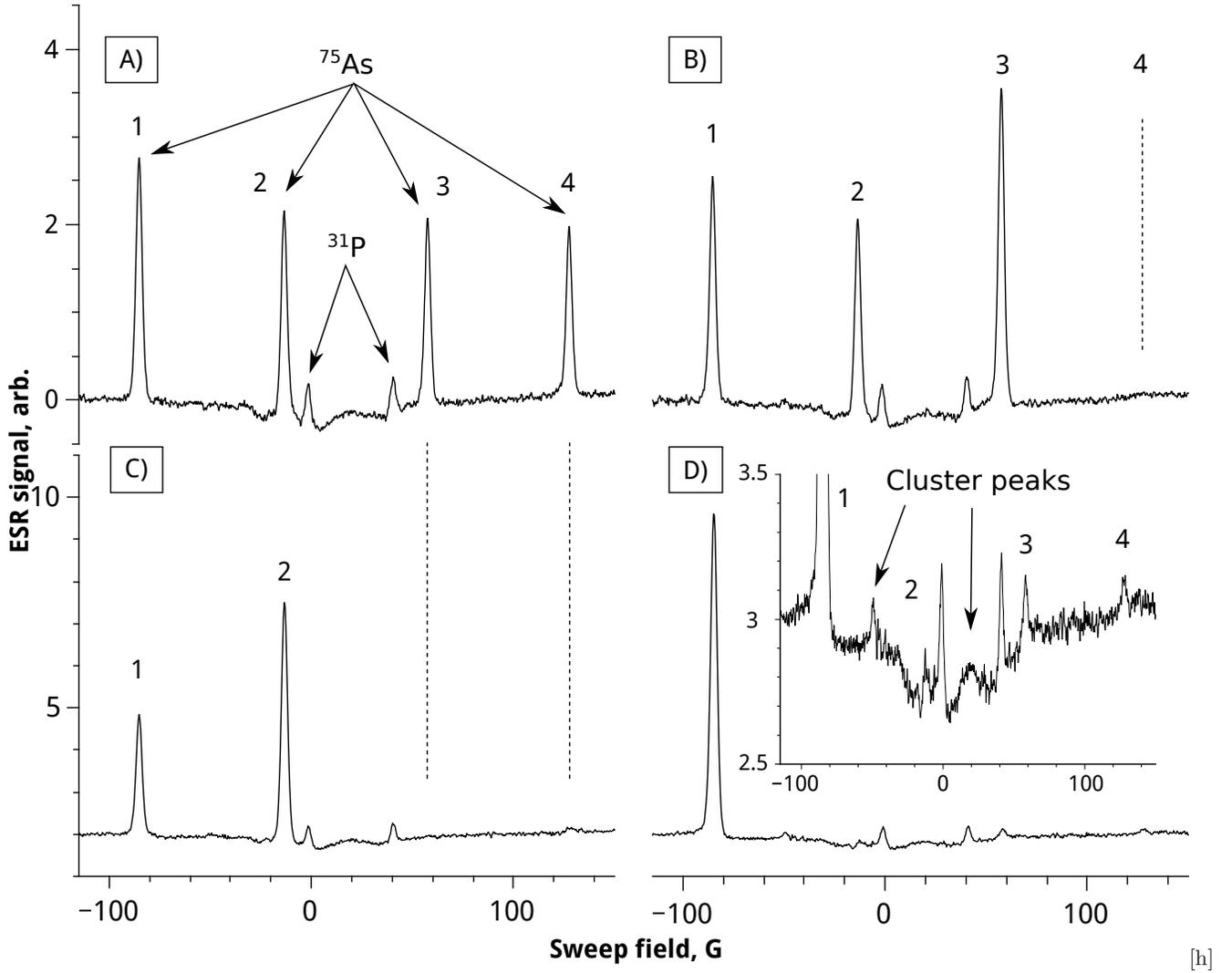}[h]
\caption{Spectrum of $^{75}$As donors and their DNP (fig. \ref{fig:Diag}) in silicon. A) Spectrum at thermal equilibrium. The four As lines are visible together with lines of impurity $^{31}$P. B) Spectrum after exciting the 4th As line with about 20 $\mu$W mw power for 2000 s. C) Spectrum after exciting 3rd line with same power. D) Spectrum after exciting 2nd line for 1500 s. Almost all of the As nuclei has been polarized to 3/2 spin state. The inset shows magnification of the spectrum. Small As population remains in the states 2-4 due to weak flip-flip relaxation. The weak line between 1st and 2nd line is probably originating from As clusters.}
\label{fig:1:spectra} 
\end{figure}

The allowed ESR lines of \As are separated by 70.8 G and numbered from 1 to 4 in order of increasing field as demonstrated in fig. \ref{fig:1:spectra}. The detected linewidth is about 3.5 G due to inhomogeneous broadening caused by the interaction with nearby spin-1/2 $^{29}$Si nuclei. The flip-flop transitions are located 23.3 G to higher field from the allowed lines 1-3 and are too weak to be detected directly in these experiments. The relaxation via allowed transitions at low temperatures is caused by the direct phonon process\citep{Feher1959a}, which provides rather long relaxation time of about 0.1 s. Therefore, in order to avoid saturation of the ESR lines during detection, the excitation power is reduced to $<$1 pW. The spin polarization is observed directly from the ESR spectrum since the intensity of each line is proportional to the population of the corresponding nuclear spin state. These populations are not influenced by the detection at low excitation powers, unlike in the NMR measurements where the polarization is typically destroyed. Even at the lowest temperatures of this work the equilibrium populations of all four nuclear states are equal to each other, which is seen in the fig. \ref{fig:1:spectra}A. The small additional peaks in the center of the spectrum originate from $^{31}$P which is present as an impurity in the sample. Although, the intensity of these peaks is substantially lower, we were able to build DNP and measure nuclear relaxation for $^{31}$P in the same sample and compare it with \As.

\begin{figure}
  \includegraphics[width=0.96\textwidth]{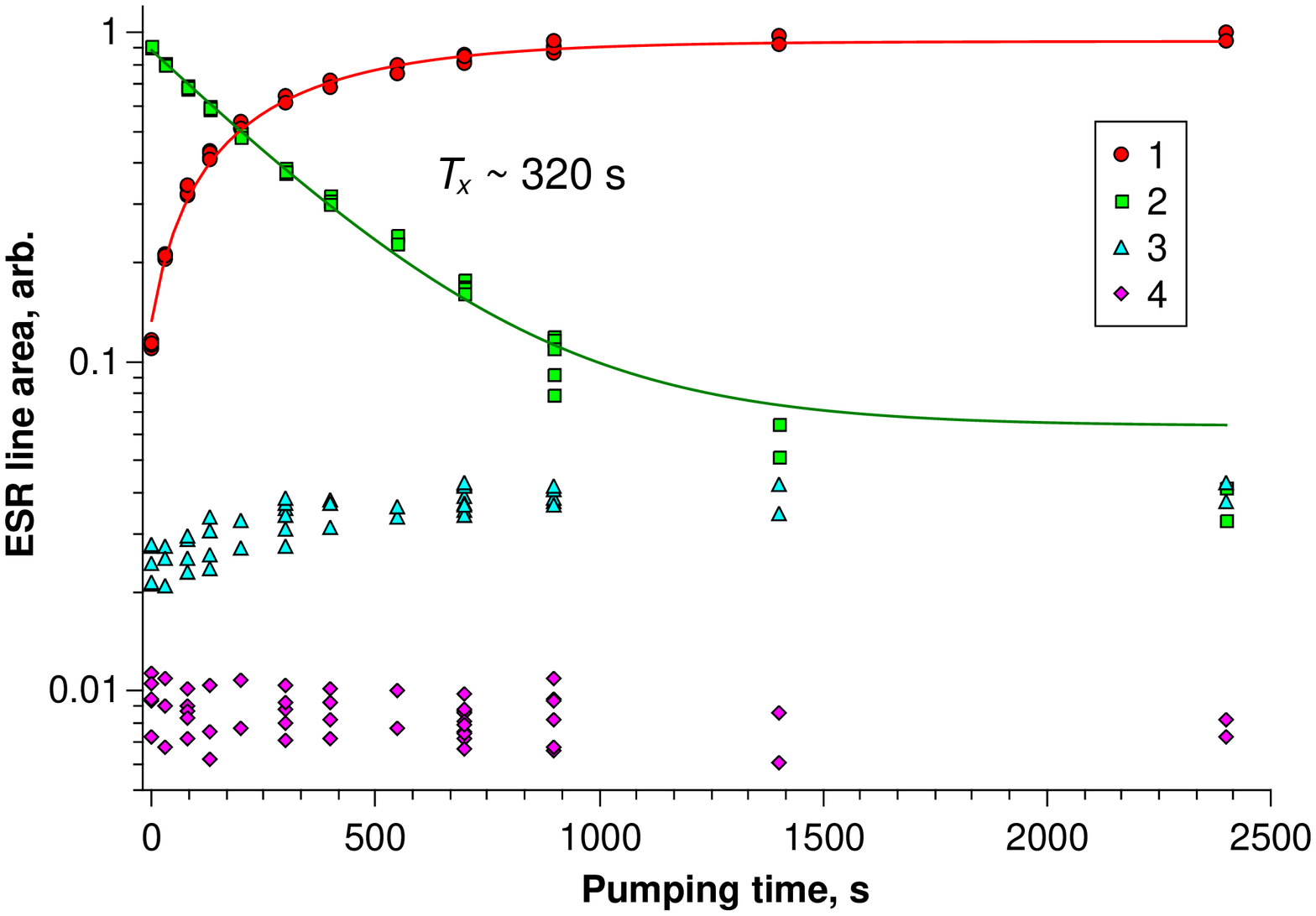}
\caption{Evolution of ESR line areas plotted as a function of cumulative pumping time when the 2nd line is pumped with FM modulated excitation at 300 mK. The line was repeatedly pumped for 20 to 1000 s time intervals and the spectra were recorded between pumping periods. The lines are fitted exponential decay and a constant giving $T_x\approx320$ s.}
\label{fig:2:dnp_evol}       
\end{figure}

Dynamic nuclear polarization is a process where the electron spin alignment is transferred to nuclei. There are several different methods of DNP \citep{Maly2008} but only Overhauser and Solid effect are relevant for \As in high magnetic fields and low temperatures.

OE DNP for \As samples is performed as follows.  The magnetic field is first tuned to match the 4th resonance transition (fig. \ref{fig:Diag}) which corresponds to the highest field line in the spectrum. The ESR excitation frequency is then modulated with 15 MHz deviation and about 10 Hz rate in order saturate the whole ESR line. The power is increased to maximum of about 20 $\mu$W and the line is "pumped" in this way for about 2000 s. Such pumping followed by the weak electron-nuclear flip-flop relaxation (fig. \ref{fig:Diag} green lines with $T_x$) transfers the population from the state $\ket{\downarrow -\frac{3}{2}}$ to the state $\ket{\downarrow -\frac{1}{2}}$ and the OE DNP is created. As a result, the area of the line 4 decreases, while that of the line 3 increases by the same amount. The pumping cycle is then repeated for 3rd line and eventually for 2nd line, finally getting all donors of the sample in the $\ket{\downarrow \frac{3}{2}}$ state. This procedure is shown in fig. \ref{fig:1:spectra}B, C and D. The flip-flop relaxation rate $1/T_x$ is much higher than flip-flip relaxation rate $1/T_{xx}$ as only a very small fraction of As spins are transferred back to higher field transitions during the 2nd and 3rd line pumping. The evolution of areas for all the four ESR lines during last cycle, pumping the 2nd line, is shown in fig. \ref{fig:2:dnp_evol}. The evolution of 1st and 2nd line fits very well with $A\exp(t/T_x)+y_0$ type function and both fits gave the same $T_x\approx320$ s.  The exponential DNP evolution differs from the $^{31}$P DNP, for which the DNP evolution has a steep start $T_x\approx20$ s and after about 300 s it slows down to $T_x\approx1100$ s.\citep{Jarvinen2014} The overall polarization time, however, is practically the same for both samples. Comparing the changes of the areas of the 1st and 3rd line due to the pumping of the 2nd line. (fig. \ref{fig:2:dnp_evol}) gives the ratio of $T_{x}/T_{xx}\approx45$ at 300 mK. The flip-flop relaxation is induced by the modulation of the isotropic hyperfine interaction while the flip-flip relaxation requires modulation of the anispotropic terms. \citep{Pines1957, Abalmassov2004} The latter is is normally much weaker, which explains the observed difference in the DNP rates.

\begin{figure}
  \includegraphics[width=0.96\textwidth]{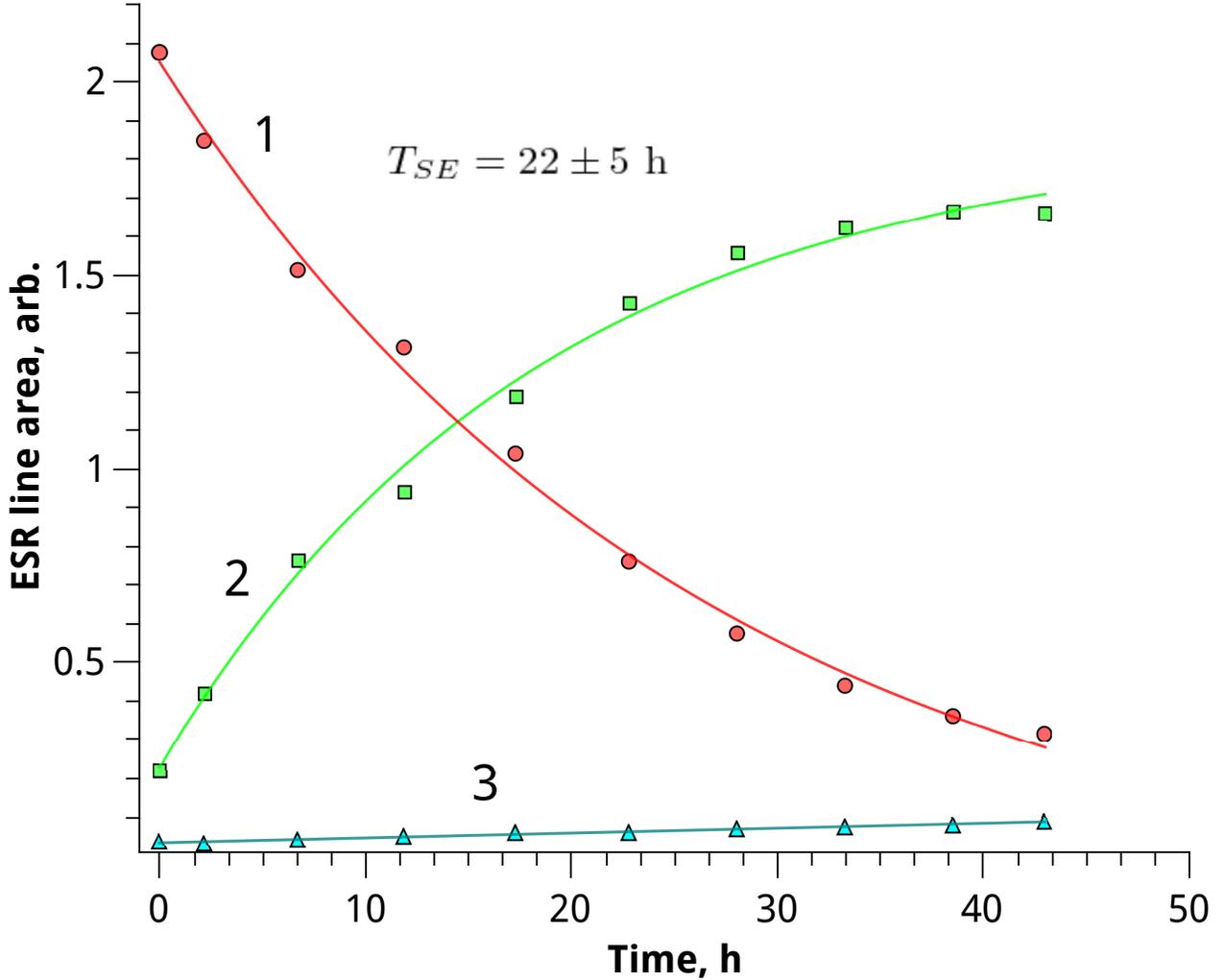}
\caption{Evolution of ESR lines 1 to 3 during solid effect pumping. The transition $\ket{\downarrow \frac{3}{2}}\longrightarrow \ket{\uparrow \frac{1}{2}}$ was pumped with 20 $\mu$W power.}
\label{fig:As_SE_evol} 
\end{figure} 

The rate of OE is highest when the allowed transition is fully saturated. Adding more power does not increase the DNP rate since the $T_x$-process is the bottleneck of the effect. Situation is different for the solid effect, where the forbidden transitions are pumped. Since their probabilities are many orders of magnitude lower, it is practically impossible to saturate them with existing sources and at low temperatures. Therefore, the efficiency of the SE DNP is typically much lower, and depends on the excitation power. The SE DNP for \As is demonstrated in fig. \ref{fig:As_SE_evol}. In the beginning sample was almost completely polarized to the $\ket{\downarrow \frac{3}{2}}$ state. Then, the $B_0$ was tuned to 23.3 G higher field from the 1st line and full power with FM modulation was turned on. The lines were measured about every two hours and pumping was continued for 44 hours. Eventually most of the area of the 1st line is moved to the 2nd. It is rather clear that 20 $\mu$W excitation power is not enough to efficiently utilize solid effect DNP. Several milliwatts of excitation power is required reach efficiency comparable with that of the OE DNP. It seems very unlikely that this could be performed without substantial overheating of the cavity and the sample. We also attempted to create SE DNP by pumping the flip-flip transitions located at 23.3 G to lower field from the 2nd line. We could not observe any effect on \As polarization after pumping for about 1000 s with full power. This is caused by even lower probability of the flip-flip transition, and the SE DNP rate is negligible in this case.

\section{Relaxation of $^{75}$As nuclei}

\begin{figure}
  \includegraphics[width=0.96\textwidth]{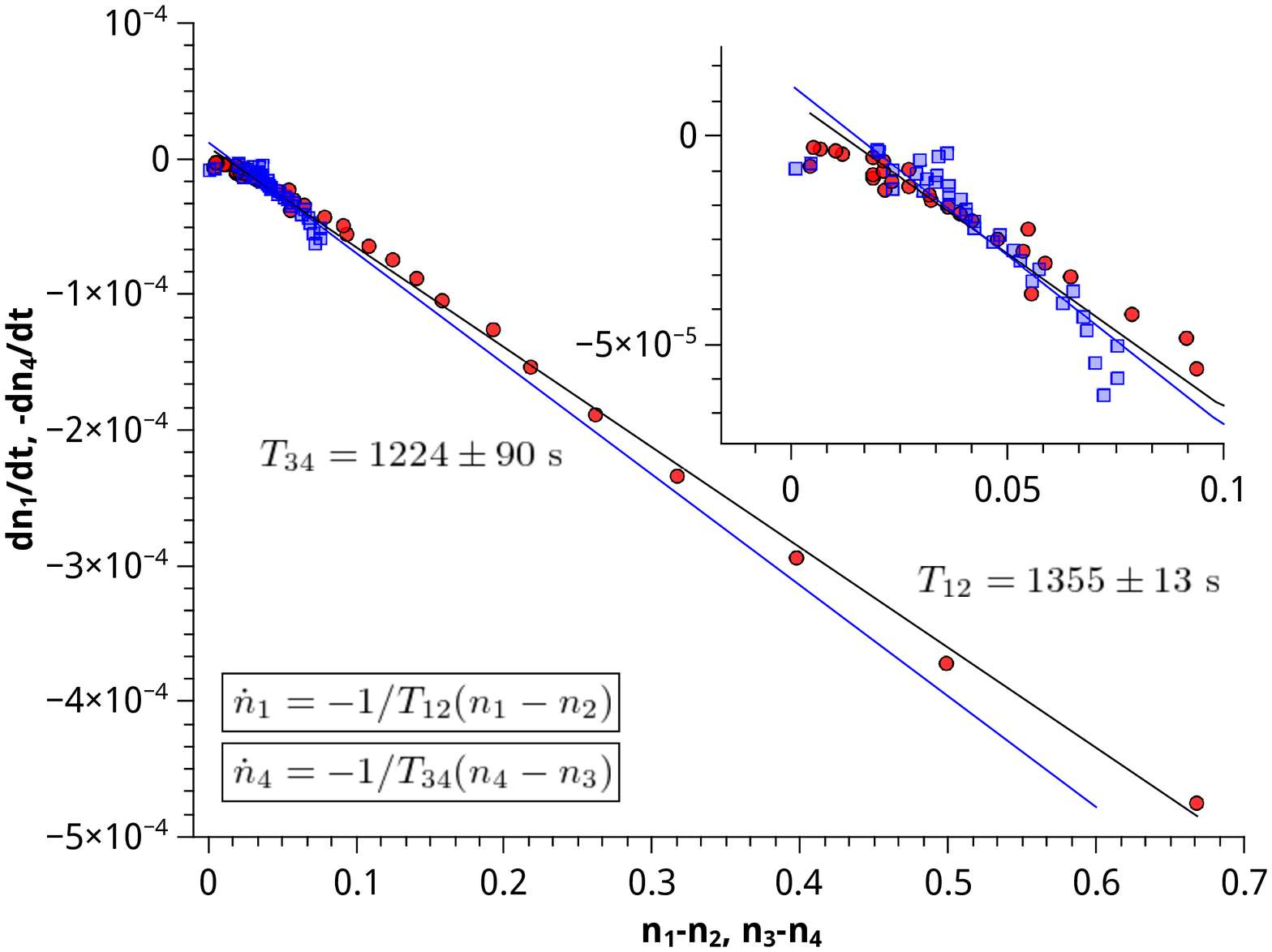}
\caption{Derivative of the nuclear state populations $\dot{n}_1$-circle(red) and -$\dot{n}_4$-square(blue) as a function of population differences. The inset shows magnifications near the zero. The relaxation times are extracted from the linear fit. The error given in the $T_{ij}$ values is root mean square error of the fit.}
\label{fig:4_Rate_area_diff} 
\end{figure}
The \As nuclear relaxation at low temperatures and high fields is expected to proceed through the Orbach process. In this process the excited electronic states are involved, and therefore, the relaxation rate has an exponential temperature dependence 
\begin{equation}\label{Orbach}
T_n\propto\exp{(\Delta/k_B T)},
\end{equation}
where $\Delta = 2 \hbar \gamma_e B_0$ is the energy separation between the upper and lower electron spin states. We performed measurements of temperature dependence of the nuclear relaxation rate  using the following procedure. First, the above described OE DNP sequence was utilized to transfer population to the state $\ket{\downarrow \frac{3}{2}}$ at temperature of 300 mK, starting with the condition of the fig. \ref{fig:1:spectra} D. It turned out that the nuclear relaxation is negligibly slow at this temperature. Then, temperature was rapidly raised to higher value and the evolution of nuclear states populations was measured periodically recording al lines in the spectrum. Finally, the relaxation rate was extracted by analysing the evolution of the four ESR line areas. 

The relaxation of nuclear spins can be described by the following set of rate equations. 
\begin{align}
\dot{n}_1&=(n_2-n_1)/T_{12}\label{eq:rate1}\\
\dot{n}_2&=(n_1-n_2)/T_{12}+(n_3-n_2)/T_{23}\label{eq:rate2}\\
\dot{n}_3&=(n_2-n_3)/T_{23}+(n_4-n_3)/T_{34}\\
\dot{n}_4&=(n_3-n_4)/T_{34}\label{eq:rate4},
\end{align}
where $n_i$ is population of nuclear state $i$ and $\dot{n}_i$ is its time derivative.
\begin{figure}
  \includegraphics[width=0.96\textwidth]{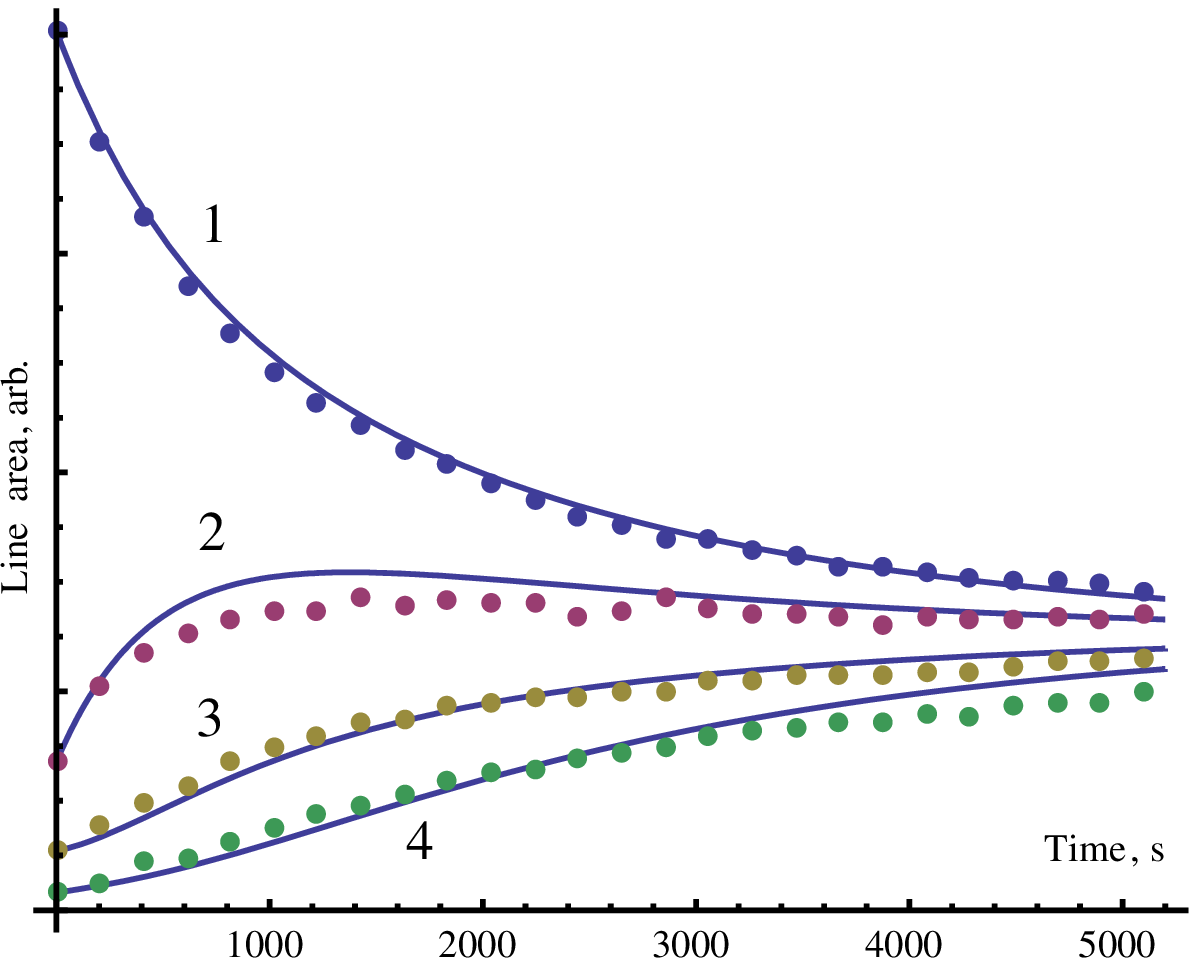}
\caption{Solutions of eq. (\ref{eq:rate1})-(\ref{eq:rate4}) fitted to ESR line area evolution at 2 K. Relaxation time $T_n=1274$ s was extracted from the fit.}
\label{fig:5_Rate_fit}       
\end{figure}
We simplified the equations by taking into account only $\Delta m_z=\pm 1$ changes of the nuclear quantum numbers.
In previous experiments with $^{31}$P doped Si the nuclear relaxation was observed to follow double exponent behavior with fast and slowly relaxing parts. If this would be the case, the eq. (\ref{eq:rate1})-(\ref{eq:rate4}) would not be adequate to describe the relaxation. To figure this out we evaluated $\dot{n}_1$ and $(n_2-n_1)$ from the evolution of ESR lines at 2 K and compared it to eq. (\ref{eq:rate1}). This is plotted in fig. \ref{fig:4_Rate_area_diff}. The dependence is linear following very well eq.(\ref{eq:rate1}) which justifies the use of eq. (\ref{eq:rate1})-(\ref{eq:rate4}) for relaxation of As spins. Also when comparing $\dot{n}_3$ to $(n_3-n_4)$ from eq. (\ref{eq:rate4}) the dependence is close to linear and the extracted $T_{34}$ is very close to $T_{12}$.

\begin{figure}
  \includegraphics[width=0.96\textwidth]{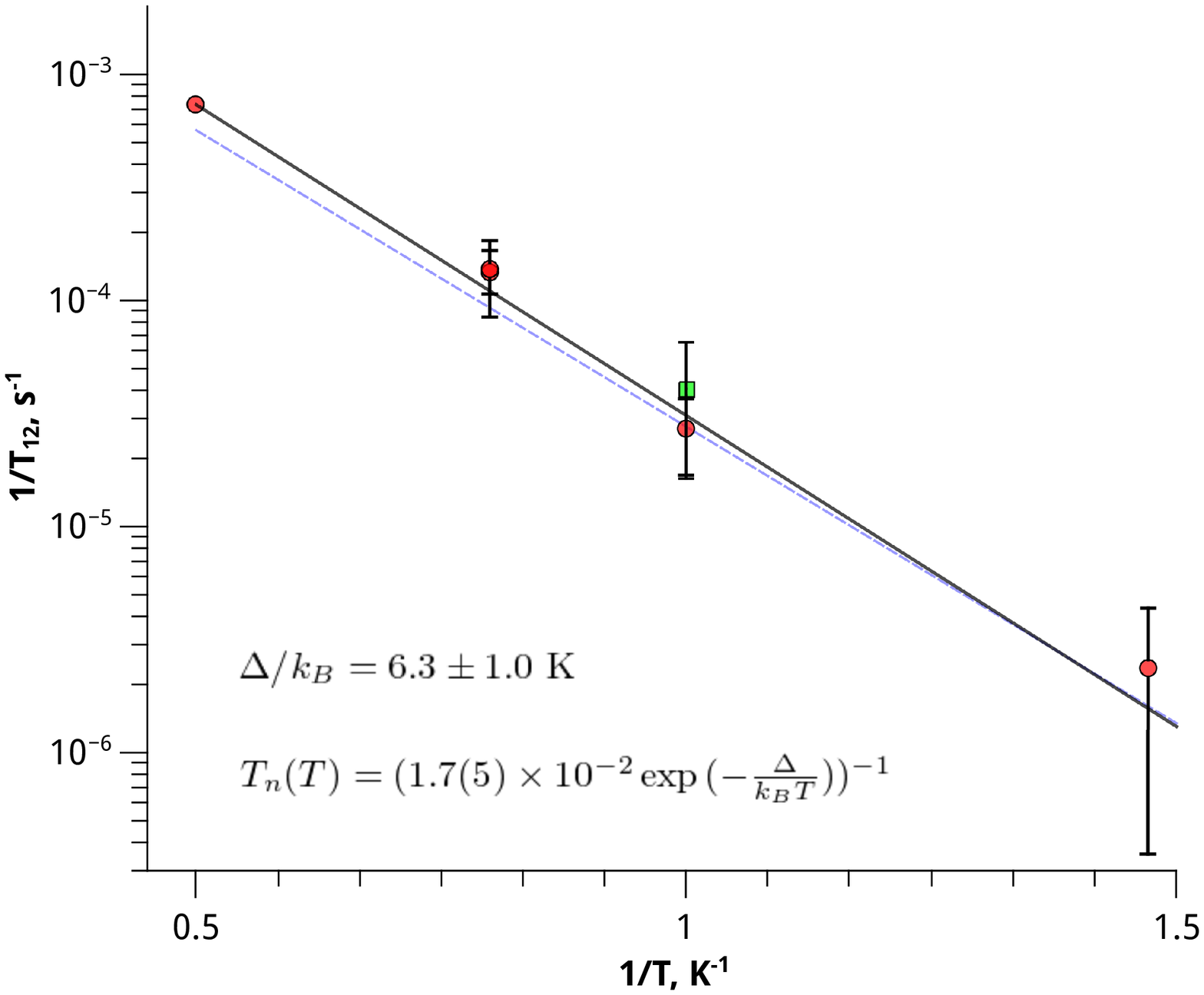}
\caption{Inverse relaxation rate as a function of inverse temperature. Solid red circles are for As and solid green square is for P dopants. Black solid line is an exponential fit to the measured values. The blue dashed line is for $^{31}$P taken from ref. \citep{Jarvinen2014}.}
\label{fig:6_Relax_temp} 
\end{figure}

At lower temperatures the evolution of the ESR line areas gets slower and extracting the derivative $\dot{n}_1$ is not so reliable. For improving the accuracy of extracting the relaxation rate the data for all four hyperfine states were fitted to the eq. (\ref{eq:rate1})-(\ref{eq:rate4}). For simplification we assumed equal relaxation times $T_n=T_{12}=T_{23}=T_{34}$. Analytical solutions for $n_1,n_2,n_3, n_4$ were calculated first with the starting conditions at $t=0$ matching the experimental data. These solutions were then fitted to the data together with only $T_n$ as a fitting parameter. Example of such a fit at 2 K is shown in fig. \ref{fig:5_Rate_fit}. The extracted $T_n=1274$ s coincides very well with the value extracted from the slope of fig. \ref{fig:4_Rate_area_diff}. The fitting was repeated for four different temperatures from 0.68 to 2 K. For the lowest temperature only $n_1$ and $n_2$ were fitted due to negligible area of the other two lines.
 
The relaxain rate $T_n^{-1}$ is plotted as a function of $T^{-1}$ in fig. \ref{fig:6_Relax_temp}. The data follow  exponential law of eq. (\ref{Orbach}) with the $\Delta/ k_b = 6.3 \pm 0.1$ K well matching the expected value in our magnetic field and thus justifying the nuclear Orbach process as the main relaxation mechanism. 

Since we have not observed bi-exponential relaxation for \As in this work, while it was found for the $^{31}$P donors in our previous experiments,\citep{Jarvinen2014} we performed similar measurements of DNP and nuclear relaxation for the small $^{31}$P impurity present in the sample studied in this work. In this case we observed single exponential relaxation, and relaxation time extracted from the fits to the experimental data well match the results for \As and for the slow relaxation process observed in our previous work. These results for the relaxation rates of $^{31}$P are also plotted in fig. \ref{fig:6_Relax_temp} for comparison.

\section{Summary}

In this work we extended our studies of the DNP in strong magnetic fields to the \As donors in silicon. Modifications of the ESR spectrometer allowed to use much larger excitation power and implement also the Solid effect DNP, which was not possible in our previous work with $^{31}$P. The nuclear relaxation was studied for \As and $^{31}$P in the same crystal and its rate was found to be the same for these donors and coincide with the $^{31}$P relaxation measured for another sample in our previous work. Now the relaxation followed single exponential law, and did no have fast component immediately after creating DNP, like it was observed earlier for $^{31}$P.\citep{Jarvinen2014} We suggested that the fast relaxation in the beginning could be caused by the influence of the $^{29}$Si spins located close to the donors. The relatively strong anisotropic hyperfine interaction of these nuclei with the electron spins of the donors may lead to the enhancement of the forbidden transition probabilities and stimulate nuclear relaxation. Since this effect is strongly anisotropic, it depends on the orientation of the crystalline lattice in the external magnetic field.  The [100] orientation of the crystalline lattice was used in the present work, which is different from the [111] orientation used earlier. In the [100] case the anisotropic part of the hyperfine interaction is smaller, and the influence of the $^{29}$Si impurity vanishes. This may explain the difference between observations of this and previous work. It is important to have slow nuclear relaxation for reaching high values of DNP, and therefore the crystal orientation in magnetic field may be optimized for that. Strong temperature dependence of the nuclear relaxation rate makes beneficial conditions of low temperatures and strong magnetic fields for increasing the efficiency of DNP.



\bibliographystyle{spphys}       
\bibliography{Si_biblio}   

\end{document}